\documentclass[aps,pre,floatfix,10pt]{revtex4-1}
\usepackage{helvet,times}
\usepackage{bm,textcomp}
\usepackage{color}

\usepackage{amsmath}

\usepackage{graphicx}
\usepackage{dcolumn}
\usepackage{latexsym}
\usepackage{textgreek}

\begin{document}
	
\setcounter{page}{1} 

\title{
Strength and stability of active ligand-receptor bonds:\\ a microtubule attached to a wall by molecular motor tethers}
\author{Dipanwita Ghanti$^\ast$, Raymond W. Friddle$^\dagger$, Debashish Chowdhury$^\ast$} 
\address{$^\ast$Department of Physics, Indian
  Institute of Technology Kanpur, 208016, India ; \\$^\dagger$Sandia National Laboratories, Livermore, CA 94550, USA }

\begin{abstract}
{We develop a stochastic kinetic model of a pre-formed attachment of a mictrotuble (MT) with a cell cortex, in which the MT is tethered to the cell by a group of active motor proteins. Such an attachment is a particularly unique case of ligand-receptor bonds: The MT ligand changes its length (and thus binding sites) with time by polymerization-depolymerization kinetics, while multiple motor receptors tend to walk actively along the MT length. These processes, combined with force-mediated unbinding of the motors, result in an elaborate behavior of the MT connection to the cell cortex.  We present results for the strength and lifetime of the system through the well-established force-clamp and force-ramp protocols when external tension is applied to the MT. The simulation results reveal that the MT-cell attachment behaves as a catch-bond or slip-bond depending on system parameters. We provide analytical approximations of the lifetime and discuss implications of our results on in-vitro experiments. }
\\
{Correspondence: debch@iitk.ac.in , rwfridd@sandia.gov}


\end{abstract}
\maketitle

\markboth{Ghanti et al.}{Stability of Microtubule-Motor Attachment}
\section*{INTRODUCTION}

Chromosome segregation is the most important process during the mitosis phase of cell cycle \cite{pollardbook}. 
In eukaryotic cells, sister chromatids, that result from chromosome replication, are segregated by a 
complex multi-component machine called mitotic spindle \cite{mcintosh12,bouck08,helmke13}. 
Microtubule (MT) \cite{lawson13}, a stiff tubular filament, whose typical diameter is about 25 nm, 
forms a major component of the scaffolding of the spindle. A MT is a polar filament in the sense that its 
two ends are dissimilar; the plus end is more dynamic than the minus end. During the morphogenesis of 
the spindle \cite{petry16,kapoor17},
MTs form transient molecular joints (non-covant bonds) with specific partners. The attachments of the 
MTs with the chromosomes, mediated by a proteineous complex called kinetochore \cite{cheeseman08,dewulf09}, 
has been under intense investigation in recent years. The forces exerted by these MTs on the kinetochores eventually pull the two sister chromatids apart in the late stages of mitosis thereby driving their journeys 
to the opposite poles of the spindle \cite{asbury17,scholey16}. 
In contrast, another set of MTs, called astral MT, diverge from the spindle poles and their distal ends 
(the so-called plus ends) form contacts with the cell cortex. The kinetochore-MT attachment has been 
under intense investigation in recent years for the obviously important role it plays in timely and accurate 
segregation of the chromosomes. However, in this paper we focus exclusively on the MT-cortex attachment.

Understanding the physics of the attachment formed by a single MT with the cortex is the first step 
in ultimately understanding how forces generated by all such attachments collectively determine the 
position and orientation of the spindle \cite{laan12,kotak13,lu13,mcnally13,pietro16}. 
The studies reported in this paper are important also from the perspective of research on MTs 
and cytoskeletal motor proteins that use MT as the `track' \cite{chowdhury13a,kolomeisky15}, 
particularly those which play crucial force-coupling roles by residing at the plus end \cite{tamura12,wade09,akhmanova15}. 
Moreover, in spite of the simplicity of the system of our study where only a single MT is attached to 
the cortex by active tethers, our analysis reveals the cooperative effects of multiple motors that give 
rise to the emergent collective properties of the attachment. Such collective phenomena are of general 
interest in several branches of physics and biology \cite{mclaughlin16}.

The main aim of this paper is to develop a minimal mathematical model of an attachment formed by a 
single astral MT with the cell cortex by capturing the essential roles of only the key components identified experimentally till now. 
The plus end of an astral MT is linked with the cortex of a living eukaryotic cell by dynein molecules 
\cite{tuncay16}. There are equispaced dynein-binding sites on the surface of a MT where the head 
of a dynein can bind specifically. Dynein is a motor protein whose natural tendency to walk towards 
the minus end of the MT track is powered by input chemical energy extracted from ATP hydrolysis 
\cite{chowdhury13a,kolomeisky15}. If the tail of a dynein is anchored on the cortex,  its attempt to 
step towards the minus end of the MT results in an effective pull on the MT directed towards the cortex.
Thus, the dyneins function as active tethers linking the MT with the cortex. The model developed in 
this paper captures only some of the key features of the real MT-cortex attachments. It is worth pointing 
out that the MT-wall model developed here differs fundamentally from another MT-wall model reported 
earlier by Sharma, Shtylla and Chowdhury \cite{sharma14} (onwards referred to as the SSC model). 
Unlike the active (i,e., energy consuming) linkers, which here mimic dynein motors, the dominant tethers 
in the SSC model are passive. Moreover, the rigid wall in the SSC model represents a kinetochore 
\cite{cheeseman08} whereas the wall in the model developed in this paper represents the cell cortex.

An attachment of the type studied in this paper can be viewed as an analog of a `ligand-receptor bond' \cite{bongrand99,karplus10} where the MT filament is the analog of a ligand while the specific binding 
partners that link its plus end with the cortex are the analogs of receptors. However, unlike common 
ligands, a MT exhibits a unique polymerization-depolymerization kinetics \cite{desai97} and the 
corresponding receptor proteins are `active' in the sense that these consume chemical fuel for their 
mechanical function. Drawing analogy with ligand-receptor bonds \cite{evans07}, we analyze the 
mathematical model to make quantitative predictions on the strength and stability of the attachment. 

We consider a pre-formed attachment to investigate its strength and stability 
using a kinetic model that mimics the protocols of dynamic force spectroscopy \cite{bizzarri12,arya16}.
Two distinct protocols are routinely used in force spectroscopy for measuring the strength and stability 
of ligand-receptor bonds \cite{bongrand99}. In the {\it force-clamp} protocol, a time-independent load tension 
is applied against the bond; the time duration after which the bond is just broken is the life time of the 
bond. Since the underlying physical process is dominated by thermal fluctuations different values of 
life time are observed upon repetition of the experiment. Therefore, the {\it stability} of the bond is 
characterized by the life time distribution (LTD). On the other hand, in the {\it force-ramp} protocol the 
magnitude of the time-dependent load tension is ramped up at a pre-decided rate till the bond just gets 
ruptured; the rupture force distribution (RFD) characterizes the {\it strength} of the bond. For a 
{\it slip-bond} the mean life time (MLT) decreases monotonically with the increasing magnitude of the 
tension in the force-clamp experiment. In contrast, a non-monotonic variation (an initial increase followed 
by decrease) of the MLT with increasing load tension in the force-clamp experiments is a characteristic 
feature of {\it catch-bonds} \cite{thomas08a,thomas08b}.

\section*{MODEL} 

Microtubules are cylindrical hollow tubes with, approximately, 25 nm diameter. 
Hetero-dimers, formed by globular proteins called $\alpha$ and $\beta$ tubulins, assemble
sequentially to form a protofilament. The length of each $\alpha$-$\beta$ dimer is about 8 nm. 
Normally 13 such protofilaments, arranged parallel to each other, form a microtubule.
However, there is a small offset of about 0.92 nm between the dimers of the neighboring 
protofilaments. In the single protofilament model \cite{anderson13} each MT is 
viewed as a single protofilament that grows helically with an effective dimer size $8/13$ nm.
Thus, we model the MT as a strictly one-dimensional stiff filament. The plus end of the MT is oriented 
along the +X-direction of the one-dimensional coordinate system chosen for the model.

In our model the cortex is represented by a rigid wall oriented perpendicular to the X-axis. 
Since a dynein motor, fueled by ATP,  has a natural tendency to walk towards the minus end of 
the MT, the molecular motors in our model are also assumed to be minus-end directed, while 
consuming input energy, in the absence of external load tension. We assume that the tail of each 
dynein motor is permanently anchored on that face of the rigid wall which faces the plus end of 
the MT. The heads of a motor, however, can either attach or detach from the MT. Each dynein 
motor has two heads that are, effectively, connected at the midpoint by a hinge. We denote the 
position of a motor by the location of its midpoint; the midpoint is assumed to be linked to the point 
of anchoring on the wall by an elastic element that is assumed to be a Hookean spring.

\begin{figure}[htb]
\center
\includegraphics[angle=0,width=0.5\textwidth]{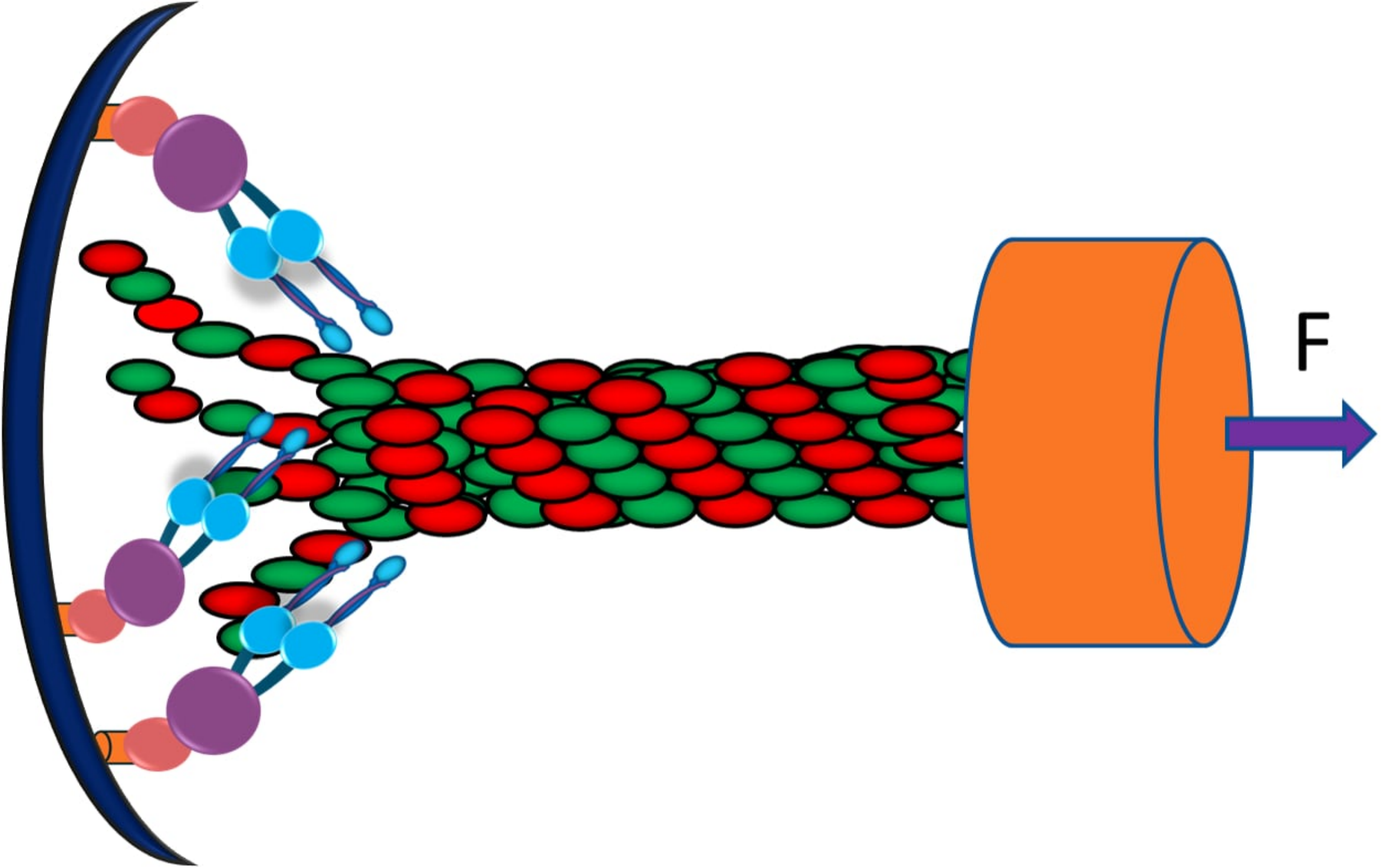}
\caption{Cartoon of a cortical dynein and MT attachment.
MT interact with cell cortex through the cortical dynein motor and dynein 
motor is attached with cell cortex (blue wall) using different protein. External 
Force $ F $ is applied on axoneme (orange) from which MTs are generated.} 
\label{fig_crtxMT}
\end{figure}

\begin{figure}[htb]
\center
\includegraphics[angle=0,width=0.5\textwidth]{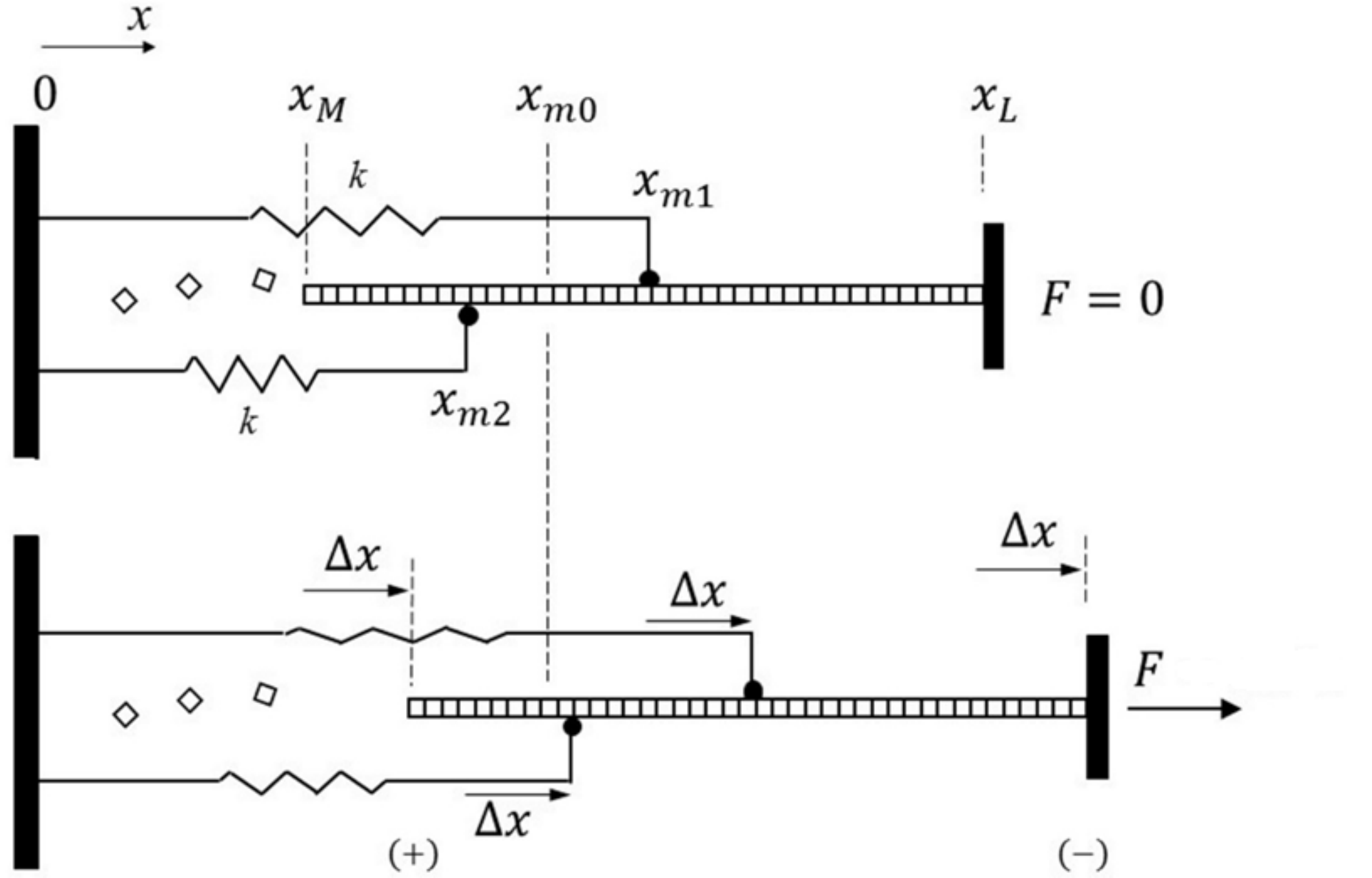}
\caption{Schematic depiction of cortical dynein MT attachment in the 
presence of external force. $ x_{M} $ represents the distance of MT tip 
from wall and mid point of the two head of two different motors are at
a distance $ x_{m1} $ and $ x_{m2} $ from the wall. 
Motors are attached with the wall through a spring with spring constant $ k $.
$ x_{m0} $ is the rest length of that spring. External force will
extend each spring by a length $ \Delta x $.} 
\label{fig_model}
\end{figure}
 

In the one-dimensional coordinate system the origin is fixed on the wall. With respect to this 
origin $x_{mi}(t)$ denotes the position of the {\it midpoint} of $ i $th molecular motor at time 
$t$ while $x_{M}(t)$ denotes the corresponding position the MT tip. 

$N_{d}$ is the total number of dynein motors that can simultaneously attach to the MT whereas $n(t)$ denotes the number of 
motors actually attached at any arbitrary instant of time (i.e., $ n(t) \leq N_{d})$. 
For any given unbound motor, ${\cal K}_{on}$ denotes the rate of binding of its head to the MT. 
Therefore,  the rate at which any unbound motor binds to the MT is \cite{muller08}
\begin{equation}
 k_{on}(n)=(N_{d}-n){\cal K}_{on}
 \label{eq_kon_d}
\end{equation}

\begin{figure}[htb]
\center
\includegraphics[angle=0,width=0.5\textwidth]{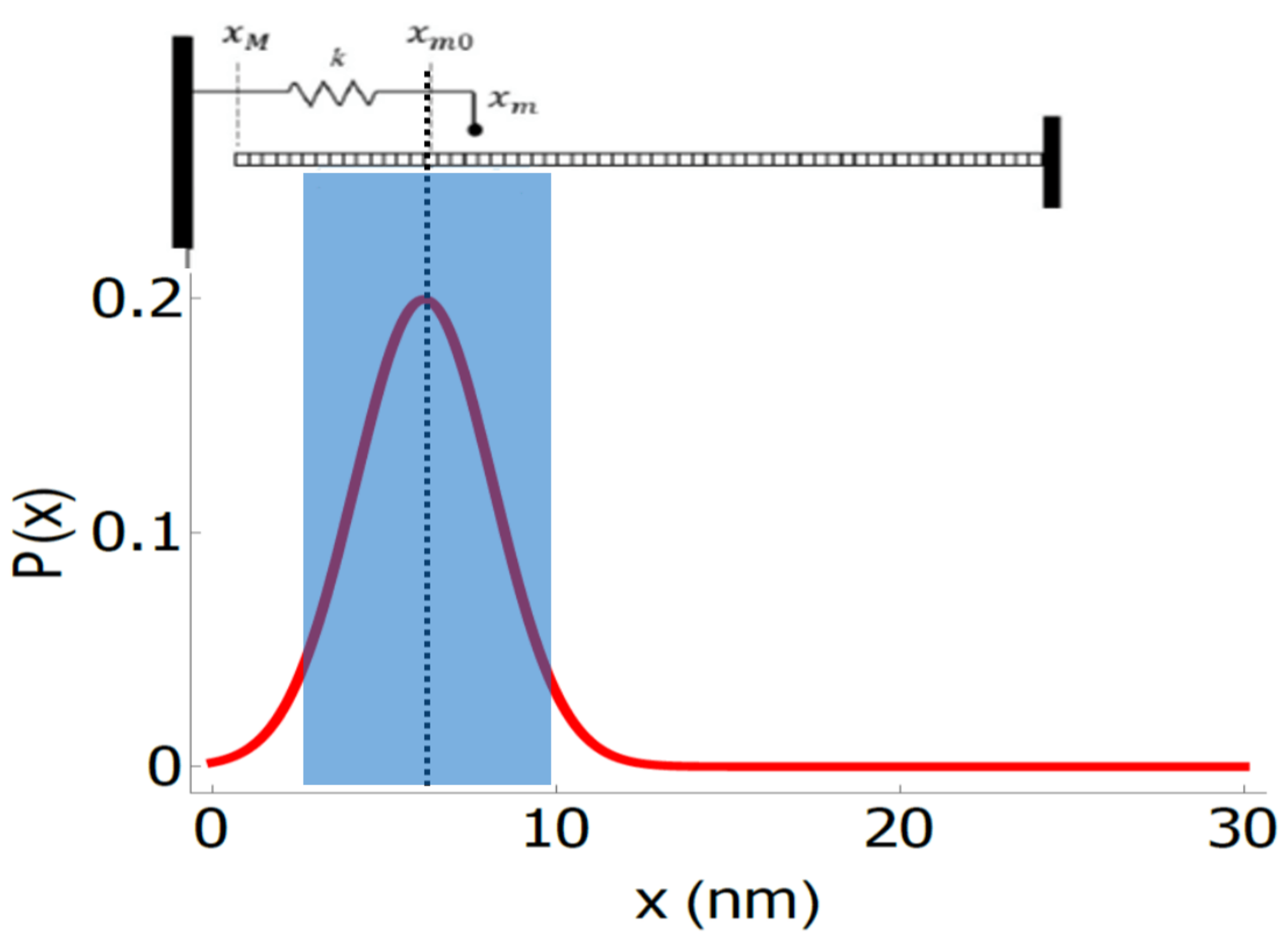}
\caption{Boltzman distribution of unbound motors connected to a wall by springs of stiffness 
$ k=1 $pN/nm and equilibrium extension $ x_{m0}=6.2 $nm, at room temperature.
Most motors will attach within the shaded area. Virtually no motors will attach beyond 15 nm.} 
\label{fig_binding}
\end{figure}

Because of being anchored on the fixed rigid wall, the motors bound to the MT cannot walk 
freely along the MT track. Note that even if a motor remains bound to a fixed site on the MT and steps neither forward 
nor backward, the spring force acting on it changes in response to the externally applied force $F$. Moreover, forward or 
backward stepping of a motor on the MT extends or compresses the spring thereby altering 
the spring force. 
 
In our model $x_{mi}$ represents the $i^{th}$ motor location ($i=1,2,...,n$).
Applying a tensile force, $F$, to the MT minus-end will move all $n$ motors by a
distance $\Delta x$ (Fig. \ref{fig_model}). 
The displacement due to external force and due to walking are contained within $x_{mi}$, and thus the force experienced by the $i^{th}$ motor is given by the corresponding force of the elastic linkage at that extension,

 \begin{equation}
 F_i = k(x_{mi}-x_{m0}) 
 \end{equation}

Following Kramers (or Bell) theory, \cite{bell78,kramers40}, we assume that the effective rate of 
unbinding one of the $n$ motors attached to the MT at a given instant of time is approximately 
\cite{muller08} 
\begin{equation}
 k_{u}(F_{i},n)=nk_{u0}e^{ { F_{i}}/F_{d}}
 \label{eq_koff}
\end{equation}

 where $k_{u0} $ denotes the rate of unbinding of the head of a single motor in the absence of load 
force. The characteristic `detachment force' $ F_{d} $ can be expressed as $ F_{d}=k_{B}T/x_{d} $ where
$ x_{d} $ is the distance from the energy minimum to maximum for the motor/MT interaction potential.


Similarly, the effective rate of forward stepping of a motor is given by
\begin{equation}
 k_{f}(F_{i})=k_{f0}e^{-{ F_{i}}\gamma/F_{sp}^{\star}}
 \label{eq_k_f}
\end{equation}

Based on experimental observations, we assume that a motor can also step towards the positive end of MT  \cite{mallik04,vale06,toba06}; 
thus we model stepping in the `reverse' direction by 
\begin{equation}
 k_{r}(F_{i})=k_{r0}e^{{ F_{i}}(1-\gamma)/F_{sp}^{\star}}
 \label{eq_k_r}
\end{equation}

\noindent where the constant parameter $ \gamma $ ($0 < \gamma < 1 $) is the fraction of the path $\ell$ over which work is done and the characteristic force 
$F_{sp}^{\star}$ can be expressed as $k_{B}T/{\ell}$ where ${\ell}$ is the length of a 
single subunit of MT. The rate $k_{r0}$ of stepping of a motor towards the plus end of the MT in the absence of load force is very small $ (k_{r0}<<k_{f0})$ because the natural direction of these motors is the minus end of MT. The form (\ref{eq_k_r}) captures the intuitive expectation that $k_{r}(F_{i})$ would increase with increasing force $ F_{i} $.


The probability that (the midpoint of) a motor is located at $x_{mi} $ and MT tip is at $ x_{M} $,
while the total number $n(t)$ of motors are bound to the MT simultaneously at that instant of 
time, is given by $P_{n}(x_{mi},x_{M},t)$. Velocity of whole MT body is given by $ v_{F} $.
Note that $x_{M}$ is a continuous variable whereas $n$ can take only non-negative integer 
values. Let $ P(\underline{x_{mi}}\vert y_{m}) $ be the conditional probability that, given a 
MT-bound motor located at site $x_{mi}$, there is another MT-bound motor at site $y_{m} $ 
on the MT $ (x_{M} < x_{mi}, y_{m}) $. Then 
$ \xi(\underline{x_{mi}}\vert y_{m}) = 1-P(\underline{x_{mi}}\vert y_{m})$
is the conditional probability that, given a motor at site $ x_{mi}$, the site $y_{m}$ is empty. 
Let $ \xi(x_{mi})$ be the probability that site $x_{mi}$ is not occupied by any motor, irrespective 
of the state of occupation of any other site. 

Under mean field approximation (MFA), the equations governing the time evolution of $P_{n}(x_{mi},x_{M}t)$  is given by 

\begin{eqnarray}
 \frac{dP_{n}(x_{mi},x_{M},t)}{dt}=
\underbrace{k_{f}(F_{i})P_{n}(x_{mi}-1,x_{M},t)\xi(\underline{x_{mi}-1}\vert x_{mi})-k_{f}(F_{i})P_{n}(x_{mi},x_{M},t)\xi(\underline{x_{mi}}\vert x_{mi}+1)}_\textrm{Forward stepping of the motor if target site is empty}
\nonumber \\
+\underbrace{k_{r}(F_{i})P_{n}(x_{mi}+1,x_{M},t)\xi(\underline{x_{mi}+1}\vert x_{mi})-k_{r}(F_{i})P_{n}(x_{mi},x_{M},t)\xi(\underline{x_{mi}}\vert x_{mi}-1)}_\textrm{Reverse stepping of the motor if target site is empty}\nonumber \\
+ \underbrace{k_{on}(n-1)(1-P_{n-1}(x_{mi},x_{M},t))-k_{on}(n)(1-P_{n}(x_{mi},x_{M},t))}_\textrm{Binding of motor to  an empty site on MT} 
\nonumber \\
+\underbrace{k_{u}(F_{i},n+1)P_{n+1}(x_{mi},x_{M},t)-k_{u}(F_{i},n)P_{n}(x_{mi},x_{M},t)}_\textrm{Unbinding of motor from an occupied site on MT}-\underbrace{v_{F}\frac{\partial P_{n}(x_{mi},x_{M},t)}{\partial x_{M}}}_\textrm{Drift velocity of whole MT body}
\label{eq_master}
\end{eqnarray}
 where 
 
\begin{equation}
  v_{F}= \frac{F-\sum_{i=1}^{n}F_i(x_{mi})}{\Gamma}
 \label{eq_velocity}
 \end{equation} 
 and  $\Gamma  $ is the viscous drag coefficient.


The rates of polymerization and de-polymerization of a MT tip are given by $ \alpha $ 
and $ \beta$, respectively. 
The rate of depolymerization of MT is suppressed by externally applied tension \cite{franck07}.
We assume that the MT-bound minus-end directed motors at the tip (plus-end) of the MT prevents 
MT protofilaments from curling outwards, thereby slowing down or speeding up depolymerization rate 
depending upon the position of the MT tip \cite{laan12}:

\begin{equation}
\beta = \beta_{0} {\rm exp}\biggl(- F_i\delta_{x_{mi},x_{M}}/F_{\star}\biggr)
\label{eq_beta}
\end{equation}		

where $F_{\star}$ is the characteristic load force at which the MT depolymerization rate is an exponentially
small fraction of $ \beta_{0} $. The Kronecker delta function ensures that the external force affects the 
depolymerization rate $ \beta $ only if the motor is bound to the tip of the MT.

The force balance equation is given by 
  \begin{equation}
  \frac{dx_{M}(t)}{dt}=\frac{F-\sum_{i=1}^{n}F_i(x_{mi})}{\Gamma}+(\beta-\alpha){\ell}+\frac{\eta(t)}{\Gamma}
  \label{eq-v_MT}
 \end{equation}
where $ \eta(t) $ is a Gaussian white noise and $ \ell$, the length of each subunit of MT is also the
 spacing between the successive motor-binding sites on the MT.

\section*{Simulation method}

\begin{table}[t]
\centering
\caption{Numerical values of the parameters used in simulation}\label{table_parameter}
\begin{tabular}{l*{2}{c}r}
Parameter             & Values\\
\hline
Spacing between binding sites on MT $ {\ell} $\cite{joglekar02,hill85,shtylla11} & $ \frac{8}{13} $  nm   \\
Rate of polymerization of MT $ \alpha $ \cite{joglekar02,hill85,shtylla11,waters96} &  30  s$^{-1} $ \\
rate of Load-free depolymerization of MT $ \beta_{0} $ \cite{joglekar02,hill85,shtylla11,waters96} & 350  s$^{-1} $ \\
Rate of Binding of motor to MT $ k_{on} $  & 3  s$^{-1} $ \\
rate of Unbinding of motor from MT $ k_{u0} $ &  3  s$^{-1} $ \\
rate of Forward stepping of motor $ k_{f0} $  & 6  s$^{-1} $ \\
rate of Backward stepping of motor $ k_{r0} $ &  0.1  s$^{-1} $ \\
Characteristic depolymerization force $ F_{\star}$ &  0.1 pN \\
Characteristic detachment force $ F_{d}$ &  1  pN  \\
Characterstic spring force of motor $ F_{sp}^{\star} $ & 1 pN  \\
Rest length of elastic linkage    $ x_{m0} $ & 6.2  nm  \\
Linkage spring constant  $ k $  & 1 pN/nm \\
Stepping parameter $ \gamma $ & 0.5 \\
Effective drag coefficient $ \Gamma $ \cite{joglekar02,hill85,shtylla11,marshall01} &  6 pNs$\mu$m$^{-1} $ \\ \hline
\end{tabular}
\end{table}

The simulations based on our theoretical model was carried out using the Gillespie algorithm 
\cite{gillespie07}. In each time step $ \Delta t $, eight types of events are possible, namely, binding/
unbinding, forward/backward hopping of any motor, polymerization and depolymerization of MT tip 
and forward/backward movement of the whole MT body.

Initially all the motors are attached to randomly selected positions on the MT, also the MT tip is placed
adjacent to the wall. Rate constants are then determined based on each motor position. A motor can unbind from its occupied site with the unbinding rate
given by eq.\ref{eq_koff}. Similarly, a motor can jump forward or backward with the transition 
rate eq.\ref{eq_k_f} and eq.\ref{eq_k_r} respectively, provided the neighboring site is empty. Finally a depolymerization event, goverened by eq.\ref{eq_beta}, can carry away a motor if it is concurrently located at the MT tip.  
 
At equilibrium, unbound motors are spatially distributed in 1D by the Boltzmann-weighted energy of their tethering springs.  (see Fig: \ref{fig_binding}) 
  
\begin{equation}
 P(x)=\sqrt{\frac{k}{2\pi k_{B}T}} \mathrm{exp}(-\frac{k(x-x_{m0})^{2}}{2k_{B}T}) 
  \label{eq-pdf}
\end{equation}
 
with $k=$ 1 pN/nm the standard deviation of unbound motor fluctuations is only $ \sigma =2 $ nm. That is, rarely would
the spring naturally stretch more than  $ \pm 2\sigma =\pm 4 $ nm.  A suitable location for binding is
 drawn from the cumulative distribution $\int_{x_{M}}^{\infty}P(x)$ by inverse transform sampling, and checked if the chosen site is empty. In this way, the position for binding is selected.
 
By the equation of motion (\ref{eq-v_MT}), the polymerization and depolymerization rates control $x_M$ by changing the length of the MT. 
But external forces influence movement of the whole MT by the resultant force acting on it, i.e. $ F-\sum_{i=1}^{n}F_i(x_{mi}) $. To treat movement of the whole MT within our discretized system, we define the corresponding rate constant by $ w=\frac{F-\sum_{i=1}^{n}F_i(x_{mi})}{\Gamma \ell} $. The sign of the expression $ w $ will decide the movement of MT in forward (positive) or backward (negative) direction.

Since the portion of available MT length is quickly depolymerized, and the instance of zero bound motors will allow the MT drift away from the wall, we neglect rebinding from the $n(t)=$ 0 state. Thus the time evolution of the motor-MT attachment is monitored until, for the first time, all the motors are detached from the MT. This first passage 
time is identified as the lifetime of the attachment for both the force clamp and force ramp conditions. We have generated trajectories of up to $10^{6} $ time steps which were then averaged to arrive at the results of interest. The common parameter values
 used in the simulation are listed in table \ref{table_parameter}.

\section*{RESULTS AND DISCUSSION}

\section*{Force clamp condition}

\begin{figure}[b]
\centering
(a)\
\includegraphics[angle=-0,width=0.45\columnwidth]{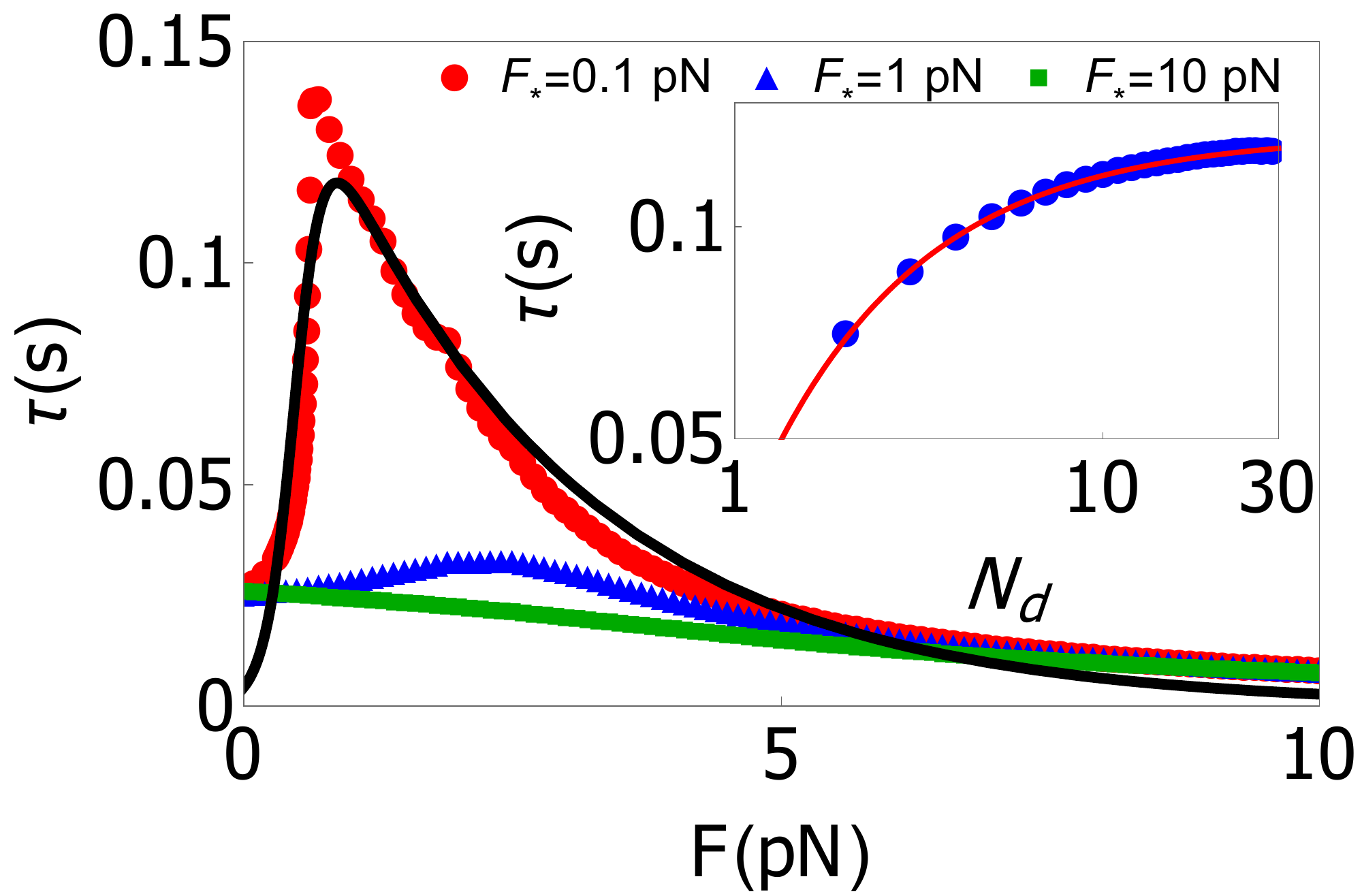}
(b)\
\includegraphics[angle=-0,width=0.45\columnwidth]{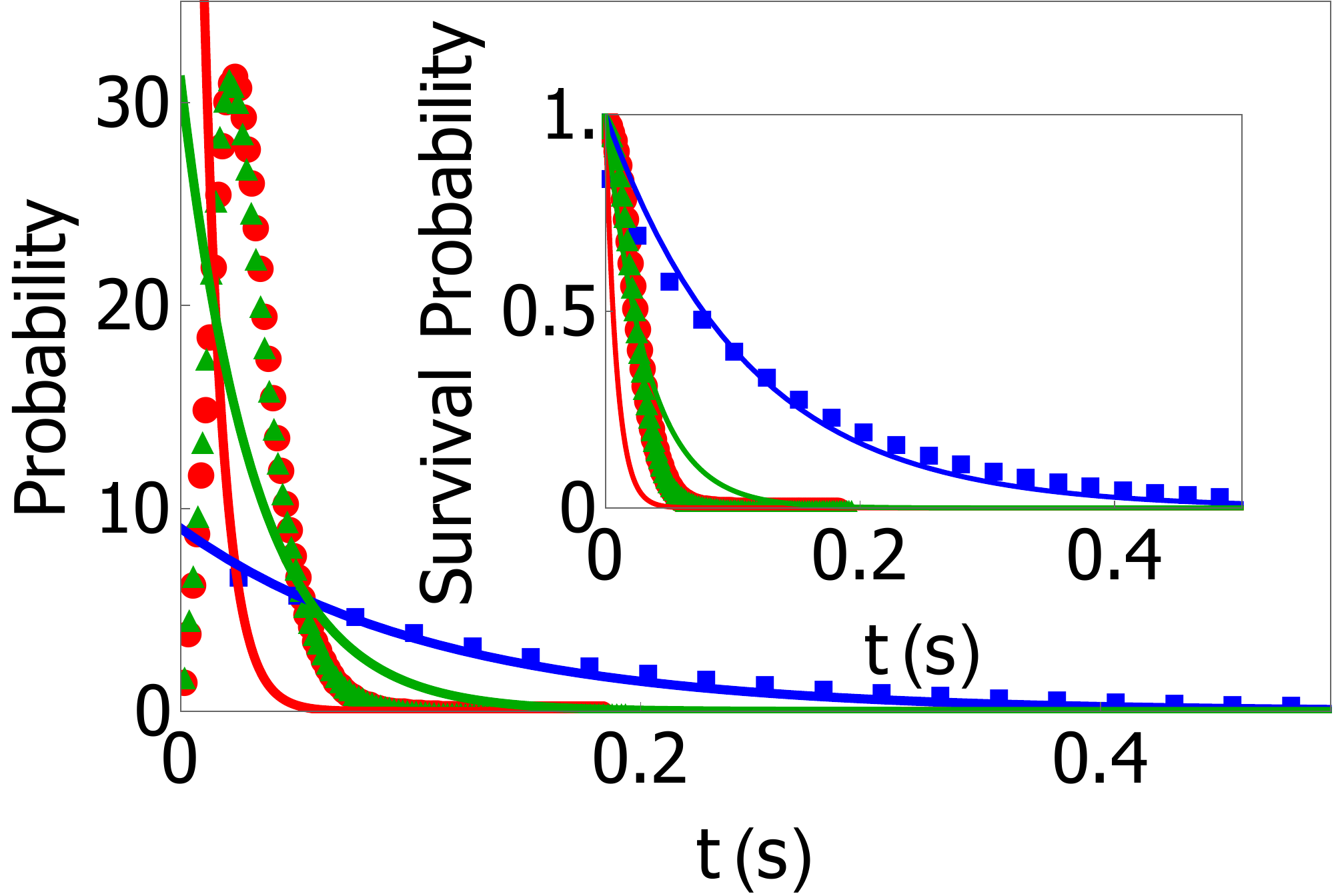}
\caption{(a) Mean lifetime $\tau$ of the attachment against the applied external
 tension $F$ for three different $ F_{\star}=0.1,1,10 $pN and fixed $N_{d}=10$. 
 The dependence of the mean life time $\tau$ on
 the number of motors $N_{d}$ is shown by the log-log plot in the inset of (a) for $ F=1 pN $ and $F_{\star}=$ 0.1 pN (solid line is Eq. (\ref{Eq_tauN})).
(b)Lifetime distribution are shown for three different external forces,
 F = 0.1pN, (red circle), F = 0.7pN (blue square) and F = 4.1pN 
(green triangle). In the inset Survival probability are shown for the same forces. }
\label{fig_dyn_clamp}
\end{figure}

Figure \ref{fig_dyn_clamp}(a) shows that the mean lifetime initially increases then decreases with external tension $ F $; such non-monotonic variation of the lifetime represents catch-bond-like behavior of cortical dynein-MT attachment. This can be explained by the two pathways which lead to removing a motor from the MT: $i$) By breaking the motor/MT bond, or $ii$) by depolymerization of a tip subunit when a motor is bound to it. 

We now to turn to finding an analytically tractable approximations of the lifetime. We coarse-grain the relevant  microscopic states of the system, $n$, into two macrostates: the completely attached ($n = N_d$) and completely detached ($n = 0$) configurations. The quantity of interest is therefore the effective transition rate (lifetime) representing a one-step process $N_d\rightarrow 0$.   

For small forces, depolymerization is faster than any other process in the system and is therefore the primary cause of motor detachment, while spontaneous unbinding adds a small contribution to the loss of motors. Neglecting rebinding and stepping dynamics for simplicity, we consider the zero-force rates of detaching all $N_d$ motors by spontaneous unbinding, $k_u^{N_d}$, or by depolymerization of motors close to the tip, $\beta^{N_d}$. In either case, decay of the bound state can be viewed as a sequence of independent Poisson processes \cite{Erdmann04,Williams03}, 
\begin{eqnarray}
k_u^{N_d} \approx \left[ \sum_{i=1}^{N_d}\frac{1}{ik_{u0}} \right]^{-1} = \frac{k_{u0}}{H_{N_d}} \\ 
\beta^{N_d} \approx \left[ \sum_{i=1}^{N_d}\frac{1}{\beta_{0}} \right]^{-1} = \frac{\beta_0}{N_d}
\label{eq_zeroRates}
\end{eqnarray} 

\noindent where $H_n$ is the $n^{th}$ harmonic number. We note that including the polymerization rate $\alpha$ adds negligible correction to $\beta^{N_d}$. The mean lifetime follows as the reciprocal of the sum of the two rates of leaving the completely attached macrostate,

\begin{equation}
\tau(F=0)\approx\left[ \frac{k_{u0}}{H_{N_d}} + \frac{\beta_0}{N_d} \right]^{-1}
\label{eq_tau_zero}
\end{equation}
\noindent For the parameters in Table \ref{table_parameter} with $N_d=10$ equation (\ref{eq_tau_zero}) evaluates to $\tau(F=0) = 27.7$ ms, which corresponds to the simulated lifetime at zero force of 27.6 ms (figure \ref{fig_dyn_clamp}a).

At large forces, the detachment of motors is no longer governed by depolymerization because the load of a motor at the MT tip will stall depolymerization according to Eq. (\ref{eq_beta}). Assuming the motors are close to one another on the MT, we approximate unbinding by the usual multiple-parallel bond treatment \cite{Seifert00,Erdmann04,Williams03} , which for $F/F_d \gg N_d$ the rate of motor removal is dictated by the time to rupture the first bond,

\begin{equation}
\tau(F\gg N_dF_d)=\frac{1}{N_dk_{u0}}\mathrm{exp}\left(-\frac{F}{N_dF_d}\right)
\label{eq_tauBigF}
\end{equation}
\noindent At $F =$ 10 pN and $N_d=10$ equation (\ref{eq_tauBigF}) gives $\tau=12.2$ ms, which coincides with the tail of all three curves at $F=$ 10 pN in figure \ref{fig_dyn_clamp}a.

In between these limits the detachment of motors is governed by one pathway that is enhanced by force ($k_u\sim\mathrm{e}^{F/F_d}$) and another that is suppressed by force ($\beta\sim\mathrm{e}^{-F/F_{\star}}$). As mentioned above we heuristically treat the intermediate force regime with effective rates of removing all motors from the fully attached state. Again neglecting the slow rebinding and motor stepping we approximate the mean detachment time as, 

\begin{equation}
	\tau(F) = \frac{1}{\kappa(F)} \approx \left[ k_1\mathrm{exp}\left(\frac{F}{N_dF_1}\right) + \frac{k_2}{N_d}\mathrm{exp}\left(-\frac{F}{F_2} \right) \right]^{-1}
 \label{eq_tauF}
\end{equation}

\noindent where the parameters of the two terms are effective parameters for the unbinding and depolymerization processes respectively. The best fit to the $ F_{\star} =$ 0.1 pN data in Fig. \ref{fig_dyn_clamp}(a) corresponds to $ k_{1}=5.55 s^{-1} $, $ N_dF_{1}\simeq 2.38 pN $ and $ k_{2}/N_d=228.5 s^{-1} $ , $ F_{2}\simeq 0.14 pN $.
For fixed force, $F =$ 1 pN, the first term of equation (\ref{eq_tauF}) depends most strongly on $N_d$, 
\begin{equation}
\tau(N_d) \approx \frac{1}{k_1}\mathrm{exp}\left(-\frac{F}{N_dF_1}\right)
\label{Eq_tauN} 
\end{equation} 
The inset of Fig:\ref{fig_dyn_clamp}(a) shows the simulated trend of lifetime with motor number. At large $N_d$ the lifetime becomes independent of $N_d$A fit of Eq. (\ref{Eq_tauN}) at $F=$ 1 pN yields $k_1 =$ 7.4 s$^{-1}$ and $F_1 =$ 0.75 pN.

We also plot the survival probability $S(t)$, the probability that till time $t$ the MT-motor 
attachment survives, in the inset of Fig:\ref{fig_dyn_clamp}(b) the survival probability of the MT-motor attachment is plotted for different values of external tension $ F $. The attachment survives longer at intermediate forces (for example, at $ F=0.7 $pN (blue square)) than at high and low forces. Each of the survival probabilities shown in Fig.\ref{fig_dyn_clamp}(b), for which $F$ has a fixed value, has been calculated using \cite{friddle12,arya16},

\begin{equation}
S(t)= {\rm exp}\biggl(-\kappa(F) t \biggr)
\label{eq_s(t)}
\end{equation}
where $\kappa(F) $ is calculated using Eq. \ref{eq_tauF}. The distributions of the corresponding life times ($-dS(t)/dt$) are plotted in Fig.\ref{fig_dyn_clamp}(b).

\section*{Force ramp condition}
\begin{figure}[t]
\centering
(a)\\
\includegraphics[angle=-0,width=0.8\textwidth]{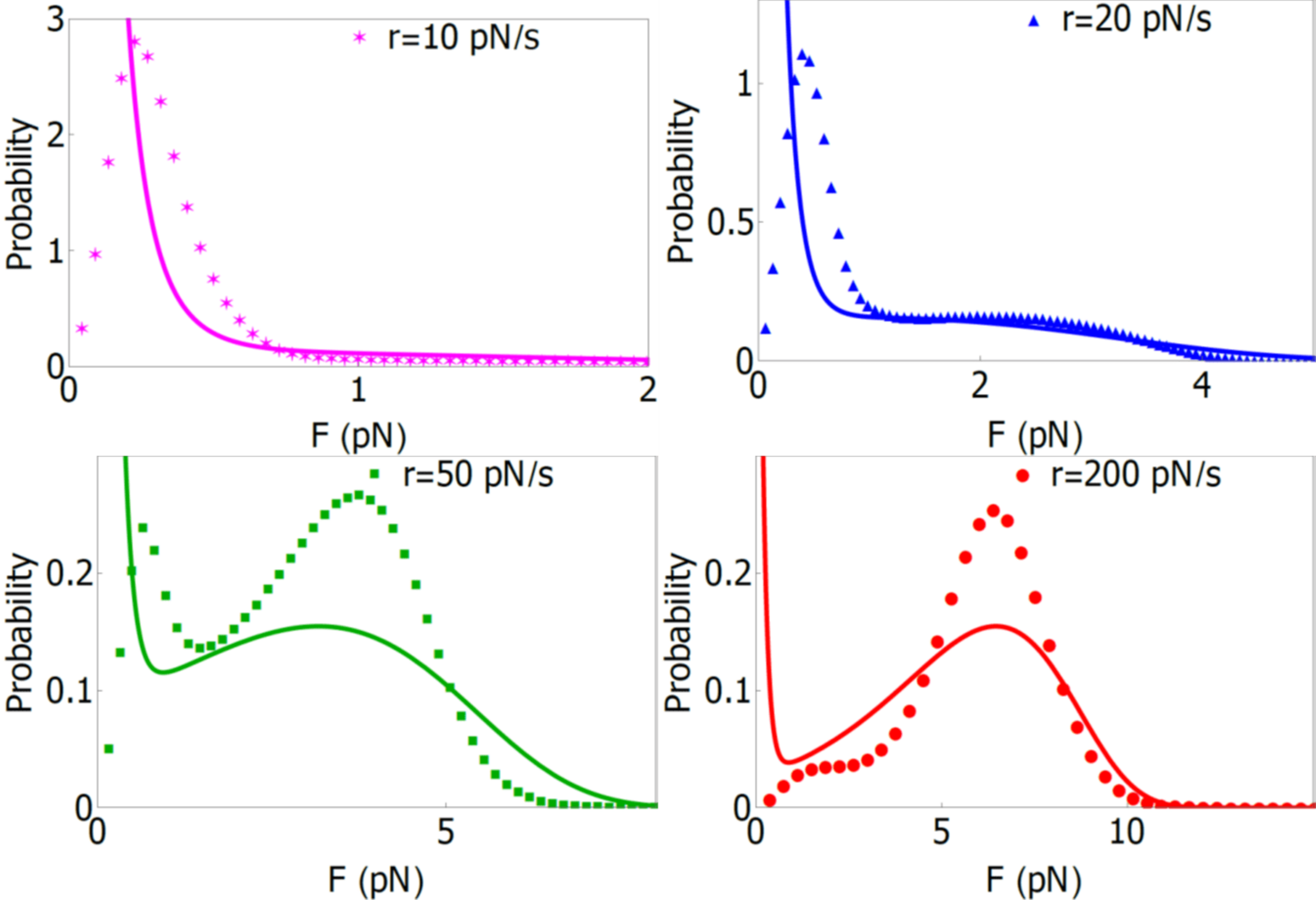}
\\(b)
\includegraphics[angle=-0,width=0.4\textwidth]{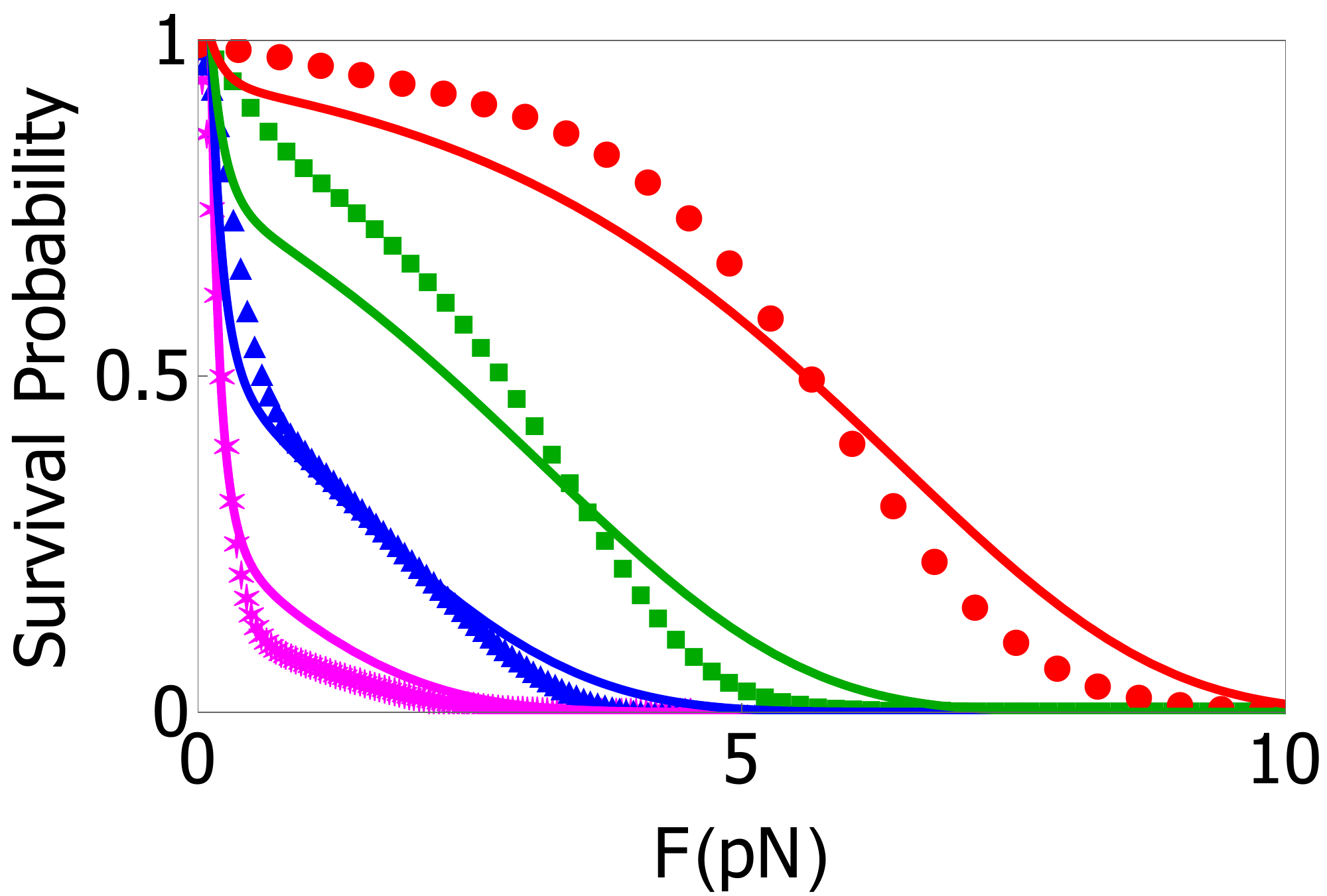}
(c)
\includegraphics[angle=-0,width=0.4\textwidth]{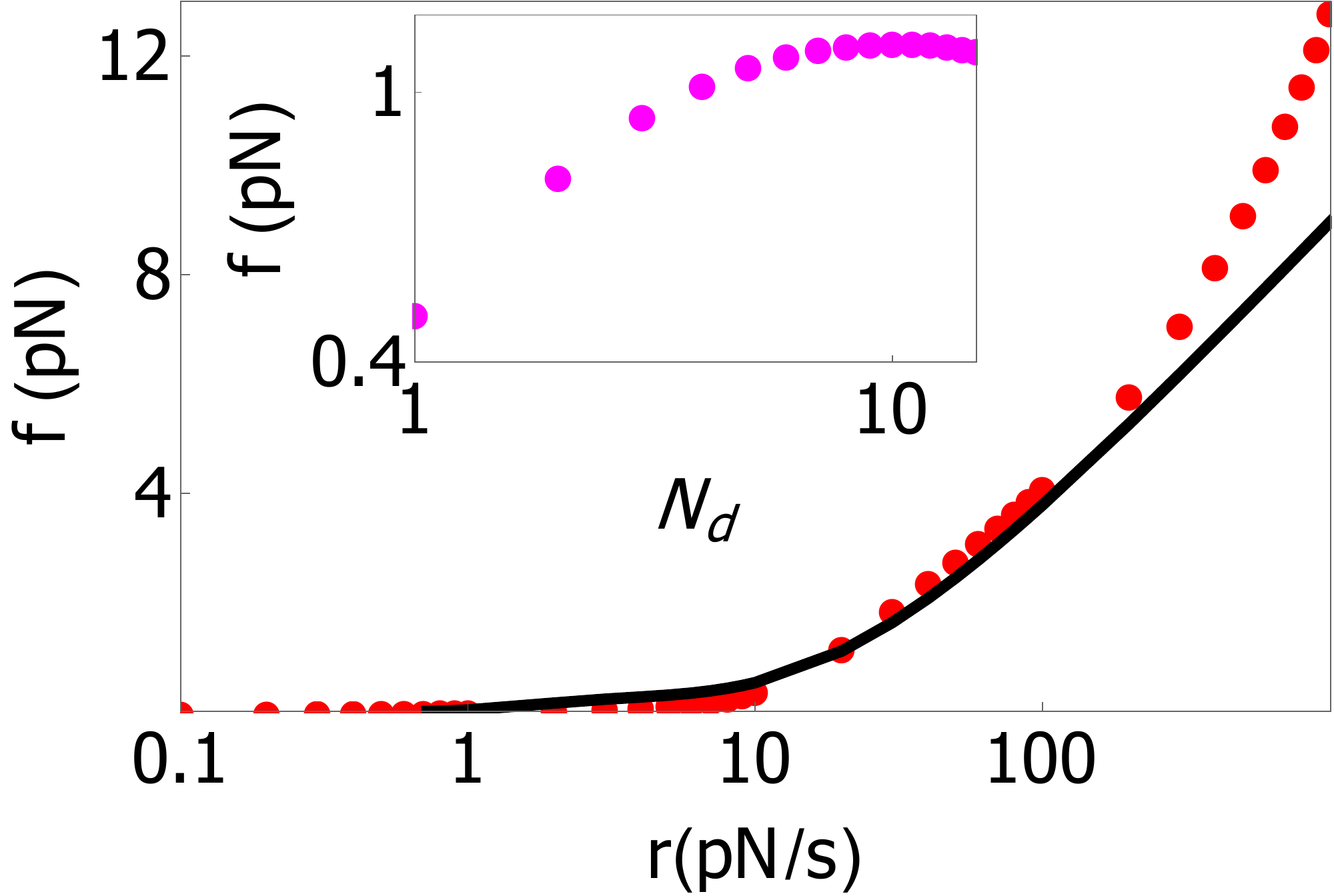}
\caption{(a) Probability density of rupture force of cortical dynein-MT attachment 
with $ N_{d}=10 $ are shown for four different loading rates, namely, $ r=10 pNs^{-1} $ 
(magenta star),$ r=20 pNs^{-1} $ (blue triangle), $ r=50 pNs^{-1} $ (green square)
and $ r=200 pNs^{-1} $ (red circle).
(b) Survival probability for different loading rates; same symbols in (a) and (b)
correspond to same set of parameters values.
(c) Mean rupture force for fixed $ N_{d}=10 $ is plotted against loading rate in a
logarithmic scale. The dependence of the mean  rupture force $f$ on the 
number of motors $N_d$ for fixed loading rate $ r=20 pNs^{-1} $ is displayed as log-log in the inset (magenta circles). 
In all the figure lines represent the best fits to the simulation 
data (please see the text for the details of the fitting procedure).  }
\label{fig_dyn_ramp}
\end{figure}

Here we present results of our simulation under the common force spectroscopy protocol of measuring the rupture force of the attachment when increasing force with time $ F(t)=rt $, and repeating over a range of loading rates, $ r $. In Fig: \ref{fig_dyn_ramp}(a) and (b) the probability distribution of rupture force $\rho(F)$  and survival probability $S(F)$ are shown for the same range of loading rates. It is important to note that at low loading rate the most probable rupture force is not zero, but instead is peaked around 0.2 pN (Fig. \ref{fig_dyn_ramp}(a), $ r=$ 10 pNs$^{-1}$). This indicates a near-equilibrium regime in which the wall-MT attachment is stabilized by the rebinding rate $k_{on}$. As loading rate increases, the rupture transitions from a near-equilbrium to a kinetic regime as a second peak emerges at high rupture force and becomes prominent (Fig. \ref{fig_dyn_ramp}(a), $ r=$ 50 pNs$^{-1}$). 

The fitting of simulation data has been done using a function \cite{friddle12,arya16}, 
\begin{equation}
S(F)={\rm exp}\biggl[-\frac{1}{r}\int_{0}^{F}\kappa(F')dF' \biggr]
\label{eq_s(F)}
\end{equation}

We compare the approximation for the lifetime in Eq.\ref{eq_tauF} against the simulated data through calculating the survival probability and corresponding probability distribution as  \cite{friddle12,arya16},

\begin{equation}
\rho(F)=-\frac{dS(F)}{dF}=\frac{\kappa(F)}{r} {\rm exp}\biggl[-\frac{1}{r}\int_{0}^{F}\kappa(F')dF' \biggr]
\label{eq_p(F)}
\end{equation}

Equation \ref{eq_p(F)} coincides with the general form of the simulated data at high forces, but diverges at low forces. This can be expected since eqs. \ref{eq_s(F)},\ref{eq_p(F)} assume a first-passage process and does not account for binding reversibility at low loading rate and low force. The mean rupture force with loading rate shows the familiar behavior of a force spectrum. At low loading rates the mean force plateaus at a very small equilibrium force, then transitions into a kinetic regime where the force goes approximately as the log of the loading rate (Fig:\ref{fig_dyn_ramp}(c)). In the irreversible approximation the mean rupture force, $f$, is given by \cite{friddle12,arya16}, 

\begin{equation}
 f  = \int_0^{\infty}F\rho(F)dF
\label{eq_Fmean}
\end{equation}

 We substitute $ \kappa(F) $ from Eq.(\ref{eq_tauF}) into  Eq.(\ref{eq_p(F)}) to calculate mean rupture force $f$ from (\ref{eq_Fmean}). We find good agreement at both low and intermediate loading rates. At large loading rates the data becomes non-linear with log-loading rate, growing to larger rupture forces than the prediction of Eq. (\ref{eq_Fmean}). This is likely explained by considering that faster loading rates allow less time for the dissociation and depolymerization processes to remove bound motors, ultimately leading to rupture of more motor-MT bonds.  

\section*{CONCLUSIONS}

Extending the earlier generalizations \cite{karplus10} of the concept of a ligand, we have 
treated a microtubule (MT) as a `ligand' that is tethered to a `receptor' wall by a group of 
minus-end directed molecular motors \cite{chowdhury13a}. The tails of the motors are 
permanently anchored on the wall while their motor heads can bind to- and unbind from 
the MT. This model of MT-wall attachment captures only a few key ingredients of the 
MT-cortex attachments in eukaryotic cells, particularly those formed during chromosome 
segregation. This minimal model incorporates the polymerization and depolymerization 
kinetics of MT. But, for the sake of simplicity, it does not include the processes of 
`catastrophe' and `rescue' that are caused by the `dynamic instability of MT filaments \cite{desai97} although these can be captured in an extended version of this model 
\cite{ghanti16}. We consider a pre-formed MT-wall attachment and carry out computer 
simulations to study statistical properties of its rupture under conditions that mimic the 
protocols of force-clamp and force-ramp experiments {\it in-vitro} \cite{bizzarri12,arya16}. 
The simulation results that we report are interpreted in the light of the theory of 
single-molecule force spectrocopy, popularized by Bell \cite{bell78} and some of its 
later generalizations \cite{arya16}. 

{\it In-vitro} experiments have been designed which remarkably resemble the conceptual model depicted in Fig.\ref{fig_crtxMT}, albeit without application of a controlled force. Laan et al. \cite{laan12} used a microfabricated vertical barrier that mimics the cell cortex. Dynein motors were anchored on the barrier and captured the MT that grew from a centrosome which was fixed on a horizontal glass surface. A slightly different experimental set up was used by Hendricks et al. \cite{hendricks12} in which a dynein coated bead was used to mimic the cell cortex. These authors demonstrated the stabilization of the MT within the broader context of the role of MT-cortex interaction in positioning of the mitotic spindle \cite{laan12,hendricks12}. However the possibility of controlling the force on the MT by an Atomic Force Microscope or Optical Trap should motivate extending these experimental setups to test the results of our theoretical model in the near future.



\section*{ACKNOWLEDGMENTS}

{DC thanks Gaurav Arya for valuable comments on an earlier preliminary version of the work. This work has been supported by a J.C. Bose National Fellowship (DC) and by the ``Prof. S. Sampath Chair'' Professorship (DC). Sandia National Laboratories is a multimission laboratory managed and operated by National Technology and Engineering Solutions of Sandia, LLC., a wholly owned subsidiary of Honeywell International, Inc., for the U.S. Department of Energy’s National Nuclear Security Administration under contract DE-NA-0003525.}\vspace*{6pt}


\end{document}